\documentclass[a4paper]{PoS}
\usepackage[utf8]{inputenc}

\title{Muon calibration of the ASTRI-Horn telescope: preliminary results}

\ShortTitle{ASTRI-Horn muon calibration}

\author{\speaker{Teresa Mineo}, Maria Concetta Maccarone, Alessio Compagnino, Milvia Capalbi, Osvaldo Catalano, Giovanni Contino, Domenico Impiombato,  Pierluca Sangiorgi\\
        INAF - Istituto di Astrofisica Spaziale e Fisica Cosmica di Palermo, via U. La Malfa 153, Palermo, Italy\\
        E-mail: \email{teresa.mineo@inaf.it}}

\author{Salvatore Garozzo, Davide Marano\\
        INAF - Osservatorio Astrofisico di Catania, Via S.Sofia 78, Catania, Italy\\
}

\author{Vito Conforti\\
        INAF - Osservatorio di Astrofisica e Scienza dello Spazio, Via P. Gobetti 93/3, Bologna, Italy\\
}

\author{for the CTA ASTRI Project\thanks{http://www.brera.inaf.it/$\sim$astri/wordpress}}

\abstract{ASTRI-Horn is a Small-Sized Telescope (SST) for very-high energy gamma-ray astronomy installed in Italy at the INAF "M.C. Fracastoro" observing station (Mt. Etna, Sicily). The ASTRI-Horn telescope is characterized by a dual-mirror optical system and a curved focal surface covered by SiPM sensors managed by a innovative fast front-end electronics.  
Dedicated studies were performed to verify the feasibility of the calibration through muons on the relatively small size of the primary mirror ($\sim$4 m diameter), as in the case of larger Cherenkov telescopes. A number of tests were performed using simulations of the atmospheric showers with the CORSIKA package and of the telescope response with a dedicated simulator.
In this contribution we present a preliminary analysis of muon events detected by ASTRI-Horn during the regular scientific data taking  performed in December 2018 and March 2019.  These muon events validate the results obtained with the simulations and definitively confirm the feasibility of calibrating the ASTRI-Horn SST telescope with muons. }

\FullConference{36th International Cosmic Ray Conference -ICRC2019-\\
		July 24th - August 1st, 2019\\
		Madison, WI, U.S.A.}

\begin{document}

\section{Introduction}
\label{sect:introduction}
Very High Energy  gamma rays (E$>$100 GeV) are completely absorbed by the Earth's atmosphere producing  showers of charged particles and  the Imaging Atmospheric Cherenkov Technique (IACT) is the most sensitive method for their observation. The complex analysis for recovering the physical characteristics of gamma rays from  images produced by showers on the focal plane of Cherenkov telescopes relies on the knowledge of the telescope optical throughput with a precision of a few percentage.  The reconstruction of the  physical and geometrical parameters of the ring-like  produced by muons  provides a simple method to achieve this request.  

When a muon hits an IACT telescope, the Cherenkov light emitted in a narrow cone along the final part of its path is imaged as a ring or arc in the focal plane camera, depending on the muon impact point onto the optical system \cite{Vacanti1994}. The position of the centre of the ring depends on the muon arrival direction respect to the telescope optics axis. The radius of the ring corresponds to the aperture of the cone (Cherenkov angle)  of the order of 1.3$^{\circ}$. The analysis of high-quality rings allows to reconstruct the muon physical parameters and to eventually extract the information useful for the calibration of the telescope optical throughput.

Muon calibration is widely used in telescopes with large collecting areas (primary mirror diameter $>$10m) \cite{Gaug2019}, but in the case of a small size telescope (diameter $\sim$4m) this method presents several challenging aspects.
A first application of muon calibration on small size telescope  has been obtained with FACT \cite{Nothe2016}, and  evaluations based on simulated data for ASTRI-Horn \cite{Pareschi2016} were presented in \cite{Mineo2016}. In this contribution we show, for the first time, the capability of muon calibration  of ASTRI-Horn using real events.

ASTRI, a small size IACT telescope developed by the Italian National Institute for Astrophysics (INAF),  and installed in Italy at the INAF ''M.C. Fracastoro'' observing station on the slopes of the Etna volcano, is named Horn in honor of  Guido Horn d'Arturo, the Italian astronomer who first proposed the use of segmented  mirrors in astronomy. 
The ASTRI-Horn telescope is characterized by a dual-mirror Schwarzschild-Couder
optical design with a 4.3-m diameter primary mirror segmented in 18 hexagonal tiles and a monolithic 1.8-m diameter secondary mirror.
The optics point spread function (PSF), defined as the 80\% of the light collected from a point like source, is contained in less than one camera pixel (0.19$^{\circ}$) over all field of view (FoV) \cite{Giro2017}.
The camera \cite{Catalano2018}, whose  sensors are Silicon Photomultipliers (SiPMs)  organized in  photon detection modules (PDMs), includes 21 units (out of 37), for a total effective FoV of 7.6$^{\circ}$. The PDMs are lodged  in a curved focal  plane and managed by a fast  electronics  system based on an innovative peak-detector technique. The camera trigger is activated when a given number of contiguous pixels, within a PDM, presents a signal above a given threshold. 

A Mini-Array composed of nine ASTRI telescopes is being developed and operated by INAF in the context of the preparatory effort for participating to CTA
\footnote{https://www.cta-observatory.org}.

The entire ASTRI-Horn system has been developed following and end-to-end approach that includes   development, deployment, operation and control of structure, mirrors, camera, inner calibration system, software and hardware for control and acquisition, data reduction and analysis software, data archiving system, and the external equipment for monitoring and calibration purposes \cite{Maccarone2017}.

\section{Analysed data sets}
\label{section:observation}
Data used for the analysis presented in this contribution are relative to the Crab observation campaign carried on by ASTRI-Horn from December 2018 to March 2019 during the telescope verification phase. They allowed us to test the sensitivity of the muon calibration to the optics degradation.

It has to be noted that during this observational campaign not all hardware components were  in nominal functional state. In particular, in the primary mirror 2 (out of 18) panels were covered, resulting in a systematic decrease of the overall amount of collected Cherenkov light. In addition, because of a strong eruption of the Etna volcano on February 2019, the secondary mirror experienced a degradation of the reflectivity and all the following observations were affected by an overall loss of the optical throughput. 
In the camera, the high-gain (HG) channel (more suitable for muon observations) was not available for technical reasons, and only data from  the low-gain (LG) channel were considered \cite{Catalano2018}. Finally, during the entire observation campaign, 1 out of  the 21  PDMs was not operative and in March 2019 a second  PDM was switched-off  because of its high level of electronic noise.  Nevertheless, the data selected for the analysis described in the following are not affected by technical problems, such as  instabilities in the signals of the PDMs. They  are relative to good atmospheric conditions and to similar altitude of the telescope. Table~\ref{tab:observation_log} shows the relevant information of the two data sets.

At the beginning of each observation, the optimal trigger configuration of the camera was evaluated with a dedicated trigger threshold scan.   In both observations  it resulted in 5 contiguous pixels above a threshold of 13 photoelectron equivalent (p.e.); the  average data acquisition rate was of the order of ~70 Hz in December and ~50 Hz in March (see Table~\ref{tab:observation_log}).

\begin{table}
\caption{Observation Log of the two data sets used in the analysis}
\label{tab:observation_log}
\begin{center}
\begin{tabular}{l|c|c}
	& \multicolumn{1}{c}{ID 1453} &  \multicolumn{1}{|c}{ID 1660} \\
\hline
Date   (yy-mm-dd)    & 2018-12-7 & 2019-3-5 \\
Duration & 23 m & 30 m \\
Pointing (J200)&RA = 46.133$^{\circ}$  & RA = 83.633$^{\circ}$  \\
                          &Dec = $+$22.014$^{\circ}$ & Dec = $+$22.015$^{\circ}$ \\
Zenith & 19$^{\circ}$~--~17$^{\circ}$  & 18$^{\circ}$~--~22$^{\circ}$   \\
Cosmic rays rate & 71  Hz & 54 Hz \\
\hline
\end{tabular}
\end{center}
\end{table}

During March 2019 observation, contemporary measurements of the diffuse night sky background (NSB) were performed through the  UVscope \cite{Maccarone2011}.
The level of the measured NSB was  of the order of  7.2$\times$10$^{12}$ ph m$^{-2}$ s$^{-1}$ sr$^{-1}$ in the wavelengths that ASTRI is sensitive to. 

Muon events were identified with preselection methods (see Sect.~\ref{sect:preselection}) and analyzed with a standard method (see Sect.~\ref{sect:ring})

\section{Flagging candidate muon ring images }
\label{sect:preselection}
Muon ring images  for the telescope calibration could be obtained through dedicated observation run under specific trigger configurations. 
On the other hand, and this is our purpose, such images can be directly collected during the regular scientific data-taking under the same trigger configuration used for gamma and protons. This implies the definition of a fast and efficient method  to select, from all the rest, possible  high-quality muon candidates. 

In the case of ASTRI-Horn, two selection methods (statistical and morphological) have been studied \cite{Maccarone2016} and properly updated with the final aim to be applied on-line at level of the ASTRI-Horn camera server where raw data registered by the camera are sent \cite{Conforti2016}. Both methods are fast with low 
computing consumption and do not need to extract those deep details which are assigned to the off-line analysis phase. 
Nevertheless, both methods need to be applied on images cleaned by the  NSB contribution. The image data are first expressed as  p.e. per pixel on which the cleaning procedure is applied. 

A very suitable and fast cleaning method, that maintains the basic shape of the signal avoiding as much as possible the presence of isolated pixels (outliers), is the two-levels cut algorithm applied on each image pixel with respect to its neighbors within a 3$\times$3 pixel window: only pixels with a signal above a given $\Phi_1$ threshold are maintained if they have at least an adjacent pixel with signal above a certain $\Phi_2$ threshold, and vice-versa. In our analysis, the values of 6 and 12 p.e. have been used for $\Phi_1$  and  $\Phi_2$, respectively. The two-level cut algorithm is then completed by the elimination of isolated outliers, if present, through a fast and iterative control of the pixels dispersion with respect to their barycenter. 
\begin{description}
\item{{\it Statistical approach.}}	
The statistical method is first based on the total size (p.e. content) of the pixels left after the cleaning. The resulting shape could be or not be a circle but it is always possible to define its barycentre, the average distance ($ave\_dista$) between pixels and barycenter, and the root mean square ($rms\_dista$) of such distance. Different threshold values applied on these parameters will allow to flag an event as possible muon candidate for calibration purposes. In the pre-selection step we  flagged events as muons candidate if they present a total size less than 500 p.e.  distributed with an  $ave\_dista$ between 2 and 7 pixels, and with $rms\_dista$ in the interval 0.5--1.5 pixel. The outcome is that more than 80\% of raw events are discarded; the remaining events are then investigated by visual inspection to be defined as ''candidate muon events''.

\item{{\it Morphological approach: the Kasa method.}}	
The peculiarity of muon events to be visualized as circular rings, closed or open, allows us to  consider a morphological approach in developing a different selection procedure to flag candidate muon events. Moreover, during this early flagging step, no deep details are required about the circular ring parameters and  a very simple algebraic method can then be applied. For this phase we adopted the Kasa method \cite{Kasa1976, Chernov2005} that uses a non-iterative procedure to fit a circle with the least algebraic distance to the set of data points (the pixels left after the cleaning).
To evaluate the shape parameters, the coordinate of the center ($xk_c$, $yk_c$) and the radius ($Rk$) are computed minimizing the function $\zeta$ (standard formula for a circle) defined as:
 
\begin{equation}
\zeta = \sum\limits_{i=1}^N((x_i-xk_c)^2+(y_i-yk_c)^2 - Rk^2)
\end{equation}

where $x_i$ and $y_i$ are the image coordinates of the $N$ pixels (points) survived after the cleaning. The algebraic distance $\zeta$ will  be zero if all points lie on a circle, and it will be small if the points lie close to a circle. 

To obtain a set of candidate muon events  for the calibration purposes, we have to apply some selection criteria. First of all, the ring images with small reconstructed Kasa radius $Rk$, i.e. small Cherenkov angle, must be rejected since these are more strongly affected by secondary ring-broadening effects that could fake the correctness of the calibration. Second, a maximum distance between the ring center and camera center has to be defined to exclude too inclined muons that are imaged at the camera edges, typically affected by inefficiencies and aberration; this corresponds to select only those rings that are fully contained in the camera. Finally, to discard too full shapes (like e.g. a proton event), we define the {\it Fullness} parameter as the ratio between the $rms$ of the reconstructed Kasa radius $Rk$ and the radius itself: in case of an empty circle, as present in a very good muon ring image, the {\it Fullness} tends to zero. In the ASTRI-Horn pre-selection step, we flagged muon rings as candidate if they have a reconstructed Kasa radius $Rk$ between 0.5$^{\circ}$ and 1.5$^{\circ}$; if their centers have a distance from the camera edge greater than or equal to the ring radius, and if they have a {\it Fullness} less than 0.26. The outcome is that more than 90\% of raw events are discarded; the remaining events are then investigated by visual inspection to be defined as ''candidate muon events''.
\end{description}

\section{Ring analysis}
\label{sect:ring}
 On the candidate muon ring images obtained by the procedures described above, a deep analysis has been performed. Each image was cleaned with the same methods and parameter values used in the preselection: the double cut with the level 6 p.e. and 12 p.e., and 
the removal of  the outliers with iterative process. To each cleaned image, the standard  analysis  \cite{Gaug2019} was applied to reconstruct the geometrical ring parameters (centre, radius and width). 
The position of the muon impact point, necessary to evaluate the number of photons hitting the primary mirror, was computed using the light distribution along  the rings. This was accumulated  from all pixels in an annulus with radius and width given by the geometrical reconstruction.

The $rms$ of the background in each pixel, due either to the electronic noise and to the NSB,  was  computed with an inverse procedure  to the one in the  image cleaning for the selection of the signals. A pixel containing the background is found if its signal is lower than 12 p.e. and there are non  adjacent  pixels containing more than 6 p.e..  This $rms$ was used to compute the errors in the signal of each pixel in addition to Poissonian errors. To take into account the variation of the NSB during the observation the $rms$ have been computed in bin of 5000 events and the appropriate value was evaluated for each event. 

\begin{description}
\item {{\it Ring geometrical parameters:}} the coordinates of the center $x_c$ and $y_c$  and the radius $R$ of the rings were obtained with a fitting procedure 
(Taubin method \cite{Taubin1991}) that minimizes the value of $\xi$ computed with the following formula:

\begin{equation}
\xi = \frac {\sum\limits_{i=1}^N((x_i-x_c)^2+(y_i-y_c)^2 - R^2)} {\sum\limits_{i=1}^N((x_i-x_c)^2+(y_i-y_c)^2)}
\end{equation}

\noindent
where $N$ is the number of pixels. 
Fig.~\ref{fig:circle} shows an example of detected ring with the relative best fit circle.

As results, December and March data sets gave the same distribution of $\xi$.  The distribution at the focal plane is not uniform as expected. However,  the most of the rings are located in three central PDMs whose efficiency resulted higher than the others.

\begin{figure}[h]
\begin{center}
\includegraphics[width=.5\textwidth]{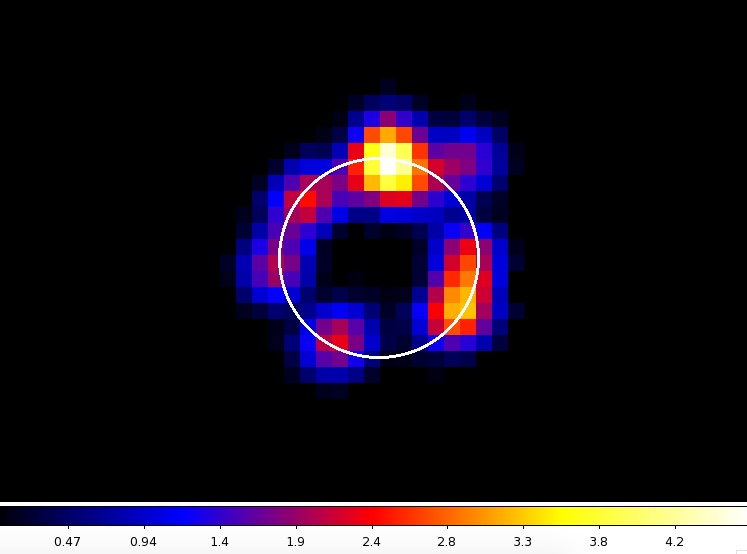}
\caption{Example of ring image in p.e., after the cleaning and a Gaussian smoothing over 3 pixels. The  best fit circle is indicated with the white circle.}
\label{fig:circle}
\end{center}
\end {figure}

\item {{\it Radial profile:}}  Since Cherenkov signal from muons  is intrinsically narrow \cite{Vacanti1994}, the width of the detected rings are then determined only by the optics PSF.  Each radial profile,  accumulated with a step of 0.5 pixel, was fitted with a Gaussian whose centre is the ring radius and whose width ($ArcWidth$) is the optics PSF. 
The  $ArcWidth$ values are consistent with the optic PSF measured  during the telescope calibration \cite{Giro2017}, and
no variation between December and March or as a function of the off-axis angles is detected. 
The distribution of the best fit values of the PSF measured with muon reconstruction is plotted in  Fig~\ref{fig:psf}.

\begin{figure}[h]
\begin{center}
\includegraphics[width=.45\textwidth]{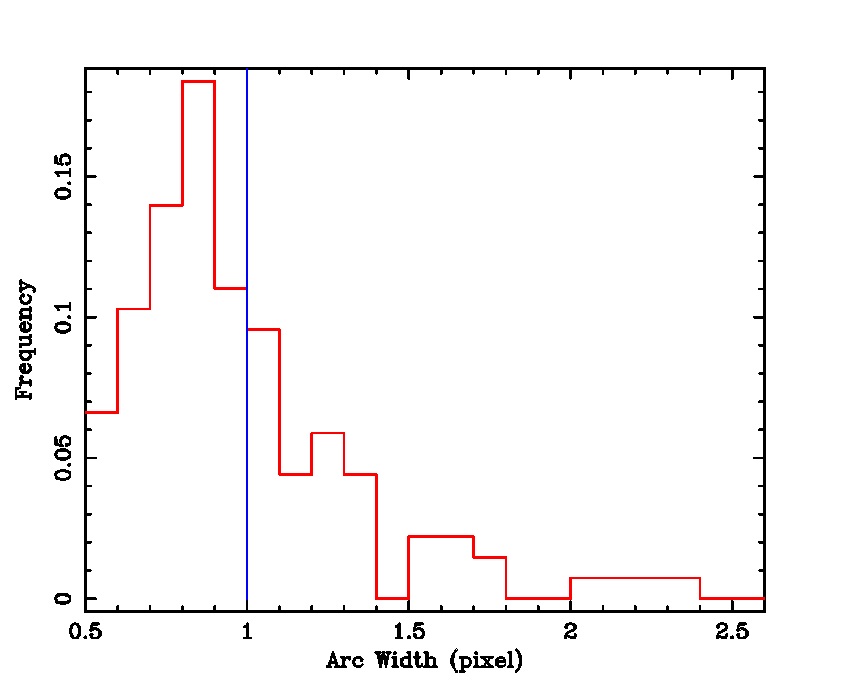}
\caption{Best fit value of the ring width fitted with a Gaussian. The  vertical blue line indicates the measured optics PSF.}
\label{fig:psf}
\end{center}
\end {figure}

\item{{\it Light distribution along the ring:}} the calibration with muons is strongly dependent on the precision on the muon impact point of the primary mirror obtained by fitting the light distribution along the ring. From the muon impact, in fact, it is possible  to evaluate the number of photons hitting the telescope \cite{Gaug2019}. 
In our analysis both mirrors are assumed with circular shapes, that allowed us to use theoretical formulas;  the  secondary mirror shadow was computed taking  into account the angle with the telescope axis of the incident direction of the muons \cite{Gaug2019}.
For each event,  the intensity profile along the ring was accumulated in 15 bins  and, after a smoothing,  fitted with the theoretical profile.
\end{description}

\begin{figure}[h]
\begin{center}
\includegraphics[width=.45\textwidth]{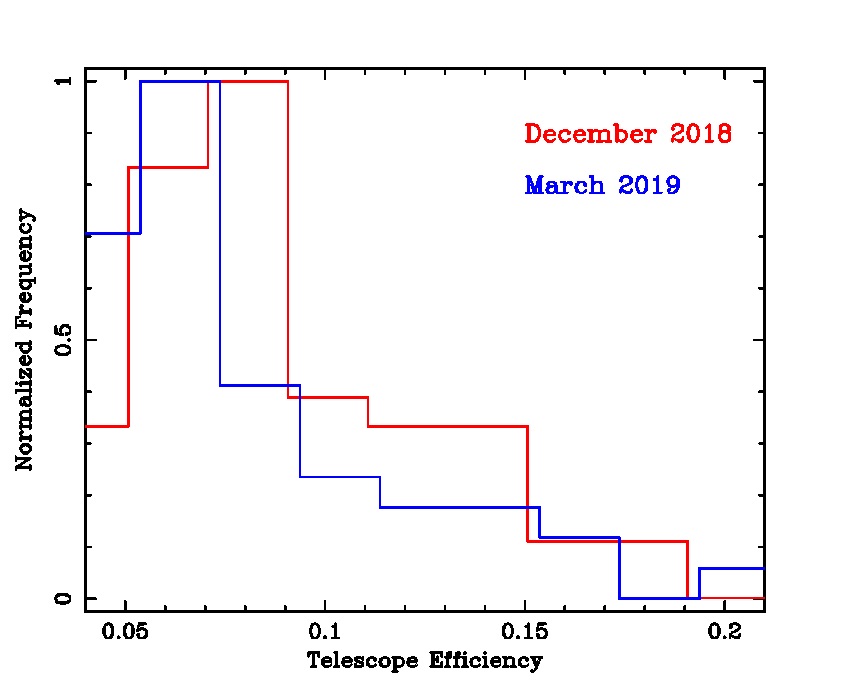}
\caption{Telescope efficiency measured with the muon analysis in December 2018 and in March 2019.}
\label{fig:efficiency}
\end{center}
\end {figure}

The distribution of  telescope efficiency, computed as the ratio between the detected p.e. and the photons hitting the primary mirror,  was obtained selecting high-quality events ($\xi\le2$ and  $\chi^2\le1$ in the fit of the light along the ring) for a robust result. We found that the efficiency peaks at ~8\% in December 2018 and decreases  to ~6\% in March 2019 as shown in Fig.~\ref{fig:efficiency}.

\section{Conclusion}
\label{sect:conclusion}
We presented a preliminary analysis of muons events detected by ASTRI-Horn during the telescope verification phase. We confirmed the significant detection of the mirror reflectivity  degradation  in the March 2019 observation after a strong Etna eruption.  
The analysis is expected to improve using all data available at the moment. 
The sensitivity of muon calibration to the optical degradation, obtained  with  ASTRI-Horn  can be definitively increased considering that the telescope was not in its nominal functioning state,  being the lack of the HG chain  the most critical failure for  this purpose.  Together with the hardware upgrades, several other improvements can be applied to the analysis as for example a more adequate cleaning method that takes into account different levels of NSB. Independent measurements of the NSB can be obtained, together with UVscope, also from the {\it variance} method \cite{Segreto2019}, not available for data used in the analysis,  that provides  pixel by pixel measurement.

\end{document}